\documentclass[%
aip,
reprint,
superscriptaddress,
ymb,
prb,
twocolumn,
titlepage,
floatfix,
showpacs,
]{revtex4-2}

\usepackage{graphicx}
\usepackage{hyperref}
\usepackage{amssymb}
\usepackage{amsmath}
\usepackage{textcomp}
\usepackage{xcolor}
\usepackage{ulem}
\allowdisplaybreaks
\usepackage{siunitx}
\definecolor{linkblue}{RGB}{49,49,148}
\hypersetup{linkcolor  = linkblue,citecolor  = linkblue,urlcolor   = linkblue,colorlinks = true,}

\makeatletter
\renewcommand*{\eqref}[1]{%
  \hyperref[{#1}]{\textup{\tagform@{\ref*{#1}}}}%
}
\makeatother

\begin{document}

\preprint{APS/123-QED}

\title{Ultrafast energy-dispersive soft-x-ray diffraction in the water window with a laser-driven source}%

\author{J.~Jarecki}
\affiliation{%
 Max-Born-Institut für Nichtlineare Optik und Kurzzeitspektroskopie, Max-Born-Straße 2A, 12489 Berlin, Germany 
}%
\author{M.~Hennecke}
\affiliation{%
 Max-Born-Institut für Nichtlineare Optik und Kurzzeitspektroskopie, Max-Born-Straße 2A, 12489 Berlin, Germany 
}%
\author{T.~Sidiropoulos}
\affiliation{%
 Max-Born-Institut für Nichtlineare Optik und Kurzzeitspektroskopie, Max-Born-Straße 2A, 12489 Berlin, Germany 
}%

\author{M.~Schnuerer}
\affiliation{%
 Max-Born-Institut für Nichtlineare Optik und Kurzzeitspektroskopie, Max-Born-Straße 2A, 12489 Berlin, Germany 
}%
\author{S.~Eisebitt}
\affiliation{%
 Max-Born-Institut für Nichtlineare Optik und Kurzzeitspektroskopie, Max-Born-Straße 2A, 12489 Berlin, Germany 
}%
 \affiliation{Technische Universität Berlin, Institut für Optik und Atomare Physik, 10623 Berlin, Germany}
\author{D.~Schick}
\email{schick@mbi-berlin.de}
\affiliation{%
 Max-Born-Institut für Nichtlineare Optik und Kurzzeitspektroskopie, Max-Born-Straße 2A, 12489 Berlin, Germany 
}%

\date{\today}

\begin{abstract}
Time-resolved soft-x-ray-diffraction experiments give access to microscopic processes in a broad range of solid-state materials by probing ultrafast dynamics of ordering phenomena.
While laboratory-based high-harmonic generation (HHG) light sources provide the required photon energies, their limited photon flux is distributed over a wide spectral range, rendering typical monochromatic diffraction schemes challenging.
Here, we present a scheme for energy-dispersive soft-x-ray diffraction with femtosecond temporal resolution and photon energies across the water window from \SIrange[]{200}{600}{eV}.
The experiment utilizes the broadband nature of the HHG emission to efficiently probe large slices in reciprocal space.
As a proof-of-concept, we study the laser-induced structural dynamics of a Mo/Si superlattice in an ultrafast, non-resonant soft-x-ray diffraction experiment.
We extract the underlying strain dynamics from the measured shift of its 1\textsuperscript{st} order superlattice Bragg peak in reciprocal space at photon energies around \qty{500}{eV} via soft-x-ray scattering simulations. 
\end{abstract}

\maketitle

\section{Introduction}

Soft-x-ray radiation is highly sensitive to a wide range of phenomena in physics, chemistry, biology, and material science.
By covering core-to-valence resonances, it can spectroscopically probe charge, spin, and orbital degrees of freedom in various material systems with element sensitivity~\cite{bres2004, fink2013}.
For solid-state research, the short wavelengths also enable few-nanometer spatial resolution, e.g., via diffraction and imaging techniques~\cite{kort2001, eise2004, chap2010, gard2017, rein2023}, and exhibit considerable penetration depth, rendering soft-x-ray techniques eligible for studying nanoscale heterostructures and buried layers~\cite{wada2009, lee2013, wies2021}.
In addition to accelerator-based installations at free-electron lasers and synchrotron-radiation facilities, laser-driven high-harmonic generation (HHG) is becoming increasingly popular as a source of short-wavelength radiation spanning from the extreme ultraviolet (XUV)~\cite{chan2009, will2015, siem2010} into the water-window, ranging from the C $K$-edge at \qty{284}{eV} up to the O $K$-edge at \qty{531}{eV} and beyond~\cite{spie1997, teic2016}.
The HHG's properties, such as spectrum, polarization, and pulse duration, are directly controlled by the driving laser and medium~\cite{huil1993, gene2019}, facilitating the application of diverse experimental techniques with a temporal resolution down to the attosecond regime~\cite{paul2001, krau2009, kret2022, kret2024}.
The key to reaching the soft-x-ray photon energy range by HHG is to increase the driving laser wavelength $\lambda\textsubscript{l}$, which at the same time leads to a severe decrease of the HHG efficiency scaling with $\lambda\textsubscript{l}\textsuperscript{-(5-6)}$~\cite{lewe1994, tate2007, shin2009}.
Furthermore, the achievable, comparatively small photon flux is also distributed over a broad spectral range, typically spanning hundreds of electron volts (eV)~\cite{schm2018, pupe2020, fu2020}.
This broadband emission renders HHG sources in the soft-x-ray range ideally suited for broadband absorption spectroscopy techniques, which have been successfully demonstrated in a series of ground-breaking experiments mainly targeting local atomic and molecular dynamics~\cite{goul2010, atta2017, smit2020, garr2022}.
Diffraction experiments, however, which can provide access to nanoscale long-range order, typically require extremely photon-inefficient monochromatization of the broad HHG spectra~\cite{fan2022}.
As a result, time-resolved diffraction experiments within the water window employing HHG sources have been elusive so far.\\
Here, we show how to benefit from the broadband and quasi-continuous nature of HHG spectra in soft-x-ray diffraction experiments by realizing an energy-dispersive scheme~\cite{Buras1968, Giessen1968} for efficiently probing large slices in reciprocal space.
We utilize femtosecond, broadband soft-x-ray pulses ranging from \SIrange{200}{600}{eV} to enable time-resolved diffraction experiments at around \qty{500}{eV} for the first time at an HHG-based setup.
For this proof-of-concept experiment, we study the photoinduced structural dynamics of a Mo/Si superlattice (SL) in non-resonant, specular diffraction geometry.
We observe a shift of the SL's 1\textsuperscript{st} order Bragg peak in reciprocal space on a few picosecond time scale.
A combination of ultrafast thermo-elastic and x-ray-scattering simulations can fully reproduce these experimental results, enabling direct and quantitative access to the underlying coherent motion of the atomic lattice.
Our findings demonstrate the feasibility and sensitivity of the energy-dispersive diffraction approach, especially for probing Bragg peak shifts in reciprocal space, independent of the HHG pulse-to-pulse intensity fluctuations.
Extending the applicability of HHG-driven soft-x-ray sources in this high photon-energy range will enable a variety of time-resolved diffraction experiments on ordering phenomena of electronic, structural, and magnetic origin~\cite{moor2016, thie2017, mitr2019, chau2020}.

\section{Results \& Discussion}

\begin{figure}[tb!]
    \centering
    \includegraphics[width=8cm]{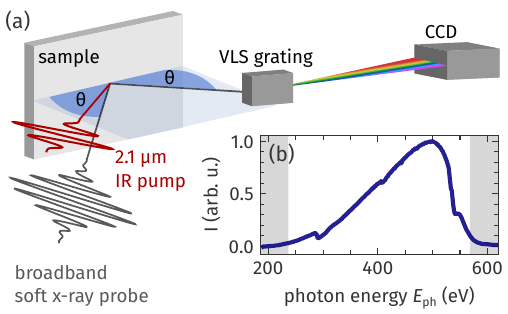}
    \caption{\label{fig:fig_1_setup}
    (a) Setup for tine-resolved energy-dispersive soft-x-ray diffraction utilizing an HHG source.
    The broadband soft-x-ray pulses hit the sample at a variable grazing angle  $\theta$ and are scattered specularly off the sample. 
    The rotatable spectrometer combines a variable-line-spacing (VLS) grating and a CCD camera to detect scattered soft-x-ray radiation with photon energy resolution.
    The \qty{2.1}{\micro\meter}-MIR pulses excite the sample quasi-collinearly with the probe pulses.
    (b) Typical soft-x-ray spectrum emitted by the HHG source ranging from \SIrange{200}{600}{eV}.
    The grey areas mark the spectral regions which are omitted in the analysis due to insufficient photon flux.
    }
\end{figure}

The general concept of our soft-x-ray diffraction experiment, combining a common $\theta/2\theta$ diffractometer with a spectrometer~\cite{henn2022}, is sketched in Fig.~\ref{fig:fig_1_setup}(a).
As HHG driver, we employ a high-average-power mid-infrared (MIR) optical parametric chirped pulse amplifier (OPCPA) system operating at a repetition rate of \qty{10}{kHz} providing pulses of \qty{27}{fs} (full width at half maximum, FWHM) duration at a central wavelength of \qty{2.1}{\micro\meter}~\cite{feng2020}.
The laser pulses are focused into a He gas cell at a backing pressure of \qty{2.3}{bar} and subsequently generate soft-x-ray pulses of $\leq \qty{27}{fs}$ FWHM duration spanning a broad spectral range from \SIrange{200}{600}{eV}.
The soft-x-ray probe pulses impinge on the sample under a grazing angle $\theta$ and get scattered into the rotatable spectrometer at an angle of $2\theta$ with respect to the incident beam. 
The spectrometer consists of a variable-line-spacing grating (VLS) and an in-vacuum CCD camera, which provide a spectral resolution of about \qty{1}{eV} at a photon energy, $E_\text{ph}$, of \qty{500}{eV}. 
A typical soft-x-ray spectrum after transmission through two Al filters of a total thickness of \qty{500}{nm} fully covering the water window is depicted in Fig.~\ref{fig:fig_1_setup}(b).
Absolute photon numbers have been determined with calibrated detection equipment previously~\cite{vanm2021}.
In the spectrum, several absorption lines corresponding to $K$-edges of residual molecules of C (\qty{285}{eV} and \qty{292}{eV}~\cite{hamo2004}), N (\qty{409.9}{eV}~\cite{teic2016}), and O (\qty{531}{eV} and \qty{538}{eV}~\cite{frat2020}), can be identified and utilized for energy calibration of the spectrometer.
A \qty{8}{\percent}-fraction of the \qty{2.1}{\micro\meter} laser light is split off and guided via a mechanical delay line to the sample for photoexcitation. 
A rotatable waveplate and polarizer set the incident fluence to \qty{9}{mJ\per cm\squared}.
Due to the non-collinearity of \qty{3.3}{\degree} between the pump and the probe pulses and their spot sizes at normal incidence of \qty{543}{\micro\meter}\,$\times$\,\qty{317}{\micro\meter} FMWH and \qty{77}{\micro\meter}\,$\times$\,\qty{111}{\micro\meter} FMWH, respectively, we achieve a temporal resolution of about \qty{40}{fs}.
Both the pump and probe pulses are $p$-polarized with respect to the scattering plane.

In a diffraction experiment probing the sample's out-of-plane (OOP) order along the $z$-direction, the magnitude of the corresponding scattering vector, $Q_z = |\Vec{Q}|=|\Vec{k}\textsubscript{out}-\Vec{k}\textsubscript{in}|$, is defined as
\begin{equation}
    Q_z = \frac{4\pi\,E_\text{ph}}{h\,c_0}\sin{\theta}\ ,
    \label{eq:qz}
\end{equation}
where $h$ denotes the Planck constant and $c_0$ is the speed of light in vacuum. 
In order to access an OOP periodicity $d$, a multiple of its corresponding reciprocal lattice vector $|\Vec{G}| = \frac{2\pi}{d}$ must match the scattering vector $\Vec{Q}$, as described by the Laue condition:
\begin{equation}
    \Vec{Q}=\Vec{k}\textsubscript{out}-\Vec{k}\textsubscript{in}=n\Vec{G} \ .
    \label{eq:Laue}
\end{equation}
Here, $\Vec{k}\textsubscript{in/out}$ denote the wavevectors of the incident and scattered soft-x-ray light, respectively, and $n$ is an integer number corresponding to the diffraction order.
Typical monochromatic diffraction schemes comprise time-consuming variations of $\theta$ and/or $E_\text{ph}$ for scanning reciprocal space, rendering themselves inconvenient and inefficient for broadband HHG sources in the water window.
The concept of probing diffraction in an energy-dispersive mode circumvents this requirement and enables access to large reciprocal-space volumes very photon- and time-efficiently in a single acquisition (see Supplemental Material).

\begin{figure}[tb!]
    \centering
    \includegraphics[width=8cm]{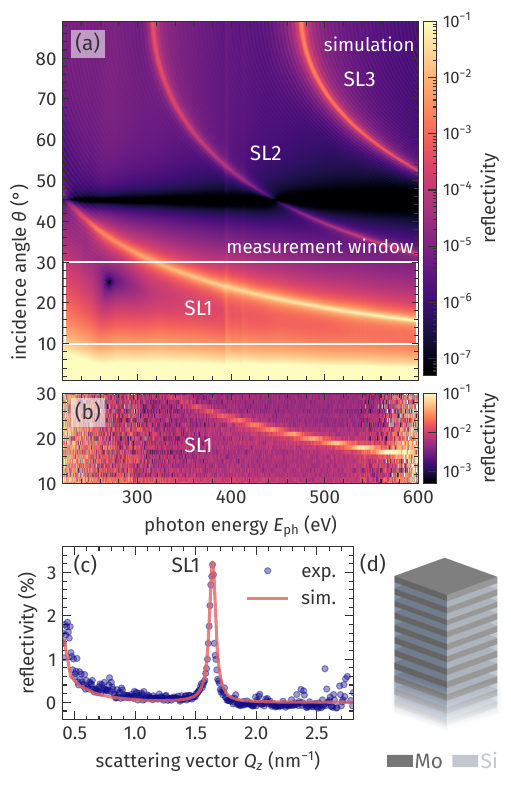}
    \caption{\label{fig:fig_2_qzscan} Energy-resolved soft-x-ray diffraction from the Mo/Si superlattice (SL) structure.
    (a)~Simulated x-ray-scattering signal of the Mo/Si SL structure [see panel (d)] with varying incidence angle $\theta$ and photon energy $E_\text{ph}$. 
    Soft-x-ray light diffracted from the periodically ordered Mo/Si layers results in distinct intensity peaks, i.e. SL Bragg peaks of order $n$ (SL$n$).
    (b)~Measurement of the diffraction signal employing broadband soft-x-ray pulses under variation of the incidence angle $\theta$ from \SIrange{10}{30}{\degree} as highlighted by the white box in panel (a).
    (c)~The simulated and experimental data from (a) and (b) are transformed into reciprocal space, revealing the 1\textsuperscript{st} order SL Bragg peak.
    }
\end{figure}

To demonstrate the capabilities of the approach and experimental realization, we study the laser-induced structural dynamics of a Mo/Si superlattice, a material system which allows to realize efficient mirrors in the XUV and soft-x-ray spectral range~\cite{stea1990, andr2002, sakh2019}.
The particular sample structure presented in this paper has already been investigated in a previous study, focusing on its use as a cross-correlator for optical laser pulses with soft-x-ray pulses at a broad range of photon energies\cite{schi2016}. 
The SL is built of 40 double layers (DL) of \qty{1.88}{nm} polycrystalline Mo and \qty{2.05}{nm} amorphous Si on a crystalline Si substrate (see Fig.~\ref{fig:fig_2_qzscan}(d)). 
The artificial unit cell of the SL is defined by the double-layer thickness $d\textsubscript{DL} = \qty{3.93}{nm}$ and the corresponding reciprocal lattice vector $G\textsubscript{DL}=2\pi/d\textsubscript{DL}$.
The structural parameters have been extracted and refined by comparing static synchrotron data with matrix-formalism-based x-ray-scattering simulations~\cite{elzo2012, schi2021} (see Supplemental Material).

We use the same simulations to model the x-ray-scattering signal from the Mo/Si SL in our energy-dispersive setup in Fig.~\ref{fig:fig_2_qzscan}(a).
We show the static result for the same angular and photon-energy range accessible in the experiment, which in principle gives access to three SL Bragg peaks of the order $n=1, 2, 3$.
While the SL Bragg peaks appear at fixed scattering vectors $\Vec{Q}$ in reciprocal space, they can be accessed by different combinations of the incidence angle $\theta$ and the photon energy $E_\text{ph}$ in real-space coordinates according to Eq.~(\ref{eq:qz}).
In Fig.~\ref{fig:fig_2_qzscan}(b), we show the experimental, energy-dispersed scattering intensity for incidence angles $\theta$ in the range of \SIrange[]{10}{30}{\degree} with a step size of \qty{1}{\degree}. 
The experimental spectra are obtained from \qty{5}{s}-integrations for each angle, resulting in a reasonably short acquisition time of the total scan.
Based on the high spectral stability of our HHG source, we can normalize the experimental data of a total angle scan to the source spectrum as shown in Fig.~\ref{fig:fig_1_setup}(b).
As a result, the experimental data is intrinsically self-normalized within every acquisition, providing accurate \textit{relative} intensities of diffraction signals.
However, because we lack an online monitor for the integrated intensity of the HHG, we do not rely on the \textit{absolute} amplitude of the diffraction signal.
Instead, we scale all acquired spectra of a scan to match the absolute intensities of the reliable x-ray-scattering simulations of the high-quality Mo/Si SL. 
This procedure does not influence the SL Bragg positions.
The measured data agrees very well with the expected shift of the SL1 Bragg peak in photon energy with varying $\theta$.
This static result already highlights the sensitivity and efficiency of the energy-dispersive diffraction approach to capture large slices in reciprocal space in a single acquisition.

For a quantitative comparison of the simulation and experimental data from panels (a) and (b) of Fig.~\ref{fig:fig_2_qzscan}, we convert both datasets into reciprocal space, following Eq. (\ref{eq:qz}). 
This transformation for all photon energies at every $\theta$ position results in a diffraction profile with a Bragg peak appearing at the same specific scattering vector $Q_z$.
To enable proper averaging of the diffraction data along the $\theta$-axis it is necessary to map the data of each scan onto the same regular $Q_z$-grid, as scanning the incidence angle leads to a shift and a different spacing of the probed reciprocal-space volume\cite{krieg2012}. 
In Fig.~\ref{fig:fig_2_qzscan}(c) we show the averaged experimental data (blue symbols) together with the simulation (red solid line) after the $Q_z$-transformation. 
In the analysis of both the experimental data and the simulation, we consider only the spectral range of sufficiently high soft-x-ray intensity in the experiment between \SIrange{230}{570}{eV} [non-shaded area in Fig.~\ref{eq:qz}(b)].
We achieve an excellent agreement between experiment and simulation by adjusting only the absolute angle of incidence $\theta$ by a small offset of \qty{0.5}{\degree} and scaling the measured intensities by a single factor for all spectra as described above.
We observe the 1\textsuperscript{st} order SL Bragg peak at $Q_z = \qty{1.626}{\per\nm}$, which is clearly shifted with respect to the predicted value following the Laue condition, c.f. Eq.~(\ref{eq:Laue}), at $Q_z^\text{Laue} = G\textsubscript{DL}=\qty{1.599}{\per\nm}$.
It has been shown, that this offset can be attributed to refraction effects introducing different phase shifts between interfering wavefronts as compared to the hard-x-ray range, resulting in the \emph{shifted} Bragg peak position~\cite{seve1998, attw2000, magn2004, wada2009}.
\begin{figure}[!b]
    \centering
    \includegraphics[width = 8cm]{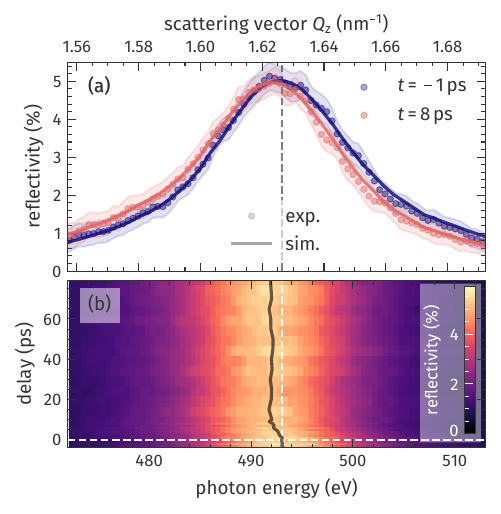}
    \caption{\label{fig:fig_3_raw_data} Transient energy-dispersive diffraction from the Mo/Si superlattice.
    (a)~The 1\textsuperscript{st} order SL Bragg peak before and after laser excitation shifts along $Q_z$ in reciprocal space or equivalently in energy in real space.
    Solid lines represent the modeled x-ray-scattering signal and circles the experimental data at a fixed angle of incidence $\theta = \qty{19}{\degree}$. The shaded area is defined by the standard error of each data point representing the uncertainty of the measurement at both delays.
    (b)~Full map of energy-dispersive reciprocal-space scans for pump-probe delays ranging from \SIrange[]{-4}{79}{ps}.
    The extracted Bragg peak position (black solid line) clearly reveals even small transient shifts in reciprocal space.
    Both panels share the scattering vector and photon-energy axis.
    }
\end{figure}
Next, we measure the laser-induced response of the Mo/Si SL in a pump-probe experiment with femtosecond temporal resolution.
The laser-driven dynamics of the Mo/Si SL can be expected to evolve as follows~\cite{schi2016}:
The laser energy is mainly absorbed by the metallic Mo layers, whereas the semiconductor Si is transparent for the \qty{2.1}{\micro\meter}-pump-laser pulses.
This results in an alternating pattern of quasi-instantaneously heated and cold SL layers with an exponentially decaying excitation amplitude along the depth of the sample.
This ultrafast heating leads to a rapid expansion of the Mo sub-lattice while the cold Si layers get compressed, leading to a complex pattern of zone-folded longitudinal acoustic phonons (ZFLAPs)~\cite{bart1999, barg2004} within the SL.
On top of this sub-ps oscillatory dynamics, a bipolar strain wave is launched from the surface of the SL towards its lower interface~{\cite{thom1986, tas1994, ruel2015}.
The picosecond time scale of this coherent strain-wave propagation is unambiguously determined by the acoustic sound velocity and thickness of the Mo/Si SL.
Simultaneously, Si also starts to expand once energy is transferred via phononic heat diffusion between the few-nanometer-thin layers of Mo and Si, resulting in an increased double-layer thickness according to the laser-induced temperature rise in both materials.
On nanosecond time scales, the SL starts to relax back to its equilibrium via heat transport to the substrate~\cite{schi2016, matt2023}.

In the transient, energy-dispersive diffraction experiment, we probe the evolution of the SL1 Bragg peak at a fixed incidence angle of $\theta = \qty{19}{\degree}$ to particularly follow the propagation of the coherent longitudinal strain wave described above.
The presented delay scan comprises 12\,loops of \qty{30}{s} integration at each delay point resulting in a total measurement duration of about \qty{4}{h}.
The experimental data, as shown in Fig.~\ref{fig:fig_3_raw_data}(a), reveal an increasing transient shift of the SL1 Bragg peak to smaller scattering vectors $Q_z$ or smaller photon energies $E_\text{ph}$.
This shift can be qualitatively understood as a direct translation of the laser-driven lattice expansion into reciprocal space via the Laue condition [Eq.~(\ref{eq:Laue})].
For the complete transient dataset, as shown in panel (b), we extract the SL1 Bragg peak position by Gaussian fits (solid black line).
To gain quantitative insights into the laser-driven lattice dynamics, we use the \textsc{udkm1Dsim} toolbox~\cite{schi2021} to simulate the thermo-elastic response of the Mo/Si SL.
To that end, we first calculate the pump laser's absorption profile according to the multi-layer absorption formalism to determine the initial spatial energy distribution.
In the second step, we use a Fourier heat-diffusion model to calculate the spatio-temporal temperature rise in the sample.
This temperature map can be used to obtain the transient strain response by solving a linear-chain model of masses and springs.
The details and parameters of these simulations are given in the Supplemental Material.
In the final step, the structural dynamics are fed as input into the same x-ray-scattering formalism, which has been benchmarked by the static HHG and synchrotron data [Fig.~\ref{fig:fig_2_qzscan}(a)] to model the transient, energy-dispersive x-ray-diffraction signal.
Using literature values for the thermo-elastic sample parameters (see Supplementary Material) and with the laser-excitation fluence as the only free parameter, we can fully simulate the shift of the SL1 Bragg peak, as shown by the solid lines in Fig.~\ref{fig:fig_3_raw_data}(a) for the two selected delays.

\begin{figure}[!ht]
    \centering
    \includegraphics[width=1\columnwidth]{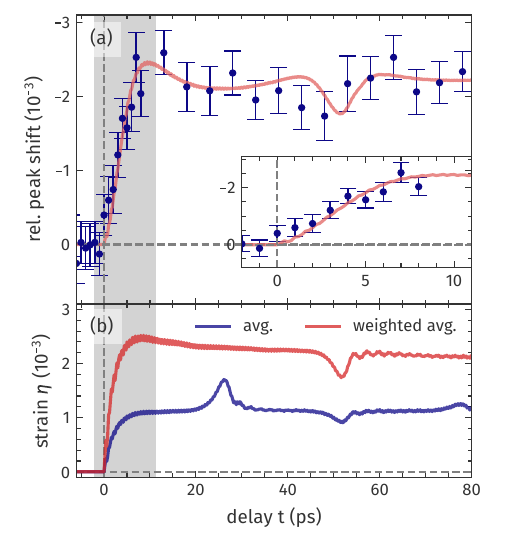}
    \caption{\label{fig:fig_4_analysis} Temporal evolution of the lattice dynamics within the Mo/Si superlattice.
    (a)~The extracted relative shift of the SL1 Bragg peak from the experimental (symbols) and simulated (solid line) x-ray diffraction.
    The inset displays a detailed view on the initial lattice dynamics marked by the shaded area.
    The error bars correspond to the statistical standard error.
    (b) Simulated average strain of the full Mo/Si SL (blue line) and weighted by the soft-x-ray absorption profile at \qty{500}{eV} (red line).
    }
\end{figure}

In Fig.~\ref{fig:fig_4_analysis}(a), we show the full temporal evolution of the relative SL1 Bragg peak position $\Delta E/E(t<\qty{0}{ps})$ extracted from both the transient experimental as well as the simulated scattering signals.
The analysis shows that the amplitude of the relative Bragg peak shift is superimposed by modulations of the transient signal, which can be attributed to the propagating strain wave generated by ultrafast laser heating.
However, comparing the transient relative peak shift to the strain averaged across the entire SL depth [blue line in Fig.~\ref{fig:fig_4_analysis}(b)] reveals substantial differences.
The modeled strain response indicates the bipolar strain wave arriving at the SL-substrate interface after \qty{26}{ps}, where it is partially reflected, resulting in a short modulation of the average lattice strain.
In contrast, the experimental peak shift reflects the relative change of the DL thickness across the probing depth of the soft-x-ray light, which is significantly shorter than the entire SL thickness of \qty{157}{nm}.
Consequently, the peak-shift signal probes only the arrival of the reflected strain wave at the surface after \qty{52}{ps} while the substrate interface at a depth of \qty{157}{nm} is not accessible.
Moreover, the observed peak shift exceeds the fitted average strain of the Mo/Si SL nearly by a factor of two.
Both effects can be understood by comparing the SL peak shift to the weighted averaged strain assuming an exponentially decaying absorption profile of the soft-x-rays [red line in Fig.~\ref{fig:fig_4_analysis}(b)]. 
We extract this probing depth to be \qty{65}{nm} and taking that information depth into account the resulting weighted strain perfectly matches the relative peak shift observed in the experiment as seen in Fig.~\ref{fig:fig_4_analysis}(a). 

\section{Conclusion}

To the best of our knowledge, we presented the first time-resolved diffraction experiment within the water window around \qty{500}{eV} photon energy at a laser-driven HHG source.
The combination of a broadband soft-x-ray source with an energy-dispersive diffractometer enables fast and efficient access to large slices in reciprocal space.
We have demonstrated the feasibility of this experimental scheme by probing the 1\textsuperscript{st} order Bragg peak of a Mo/Si SL both statically and temporally resolved after laser excitation with \qty{40}{fs} temporal resolution enabled by reasonable short acquisition times facilitating pump-probe experiments. 
We obtain excellent agreement between the static as well as transient experimental data with x-ray scattering simulations. In both experiment and theory, the underlying structural information is accessible by analyzing the corresponding Bragg peak shape in reciprocal space.
This combined experimental and theoretical approach enables an unambiguous and quantitative determination of the underlying spatio-temporal strain dynamics in the Mo/Si SL after photoexcitation on the femto- and picosecond time scale.

Based on the high stability of the spectral shape of our HHG source, we can follow Bragg peak shifts with remarkable sensitivity.
The future implementation of intensity-normalization schemes~\cite{Holtz2017, Schick2020} will enable a new class of experiments that are also sensitive to ultrafast changes of the structure factors associated with nanoscale periodicities.
Beyond this experimental proof-of-concept focused on probing the coherent acoustic lattice dynamics in a Mo/Si SL, transient and energy-dispersive diffraction experiments at HHG sources as demonstrated here will be able to target other order phenomena of spin, charge, and orbital origin.
This does require reaching the relevant core-to-valence transitions of the elements involved, in order to provide sufficient diffraction contrast.
Our state-of-the-art setup reaches the $L$-edges of the early 3d transition metals, such as Ti and V, as well as the $K$-edges of O, which already now enables a variety of studies within the broad class of oxide materials.
With new powerful laser systems at even longer driver wavelengths at the horizon, the $L$-edges of other, e.g., magnetically relevant elements such as Mn, Co, Fe, and Ni will also become within reach for energy-dispersive diffraction experiments at laser-driven HHG sources.

\begin{acknowledgments}
J.J. and D.S. would like to thank the Leibniz Association for funding through the Leibniz Junior Research Group J134/2022.
This work was carried out at the Nanomovie Application Laboratory at the Max Born Institute, which was established with the help of the European Regional Development Fund.
\end{acknowledgments}

\appendix

\section{Energy-dispersive diffraction with a white-light source}
In this section, we illustrate the concept of energy-dispersive diffraction compared to common monochromatic diffraction schemes.
In a diffraction experiment probing the sample's out-of-plane (OOP) order along the $z$-direction, the magnitude of the probed scattering vector, defined as the difference between the wave vectors of the incident and scattered soft-x-ray light $Q_z = |\Vec{Q}|=|\Vec{k}\textsubscript{out}-\Vec{k}\textsubscript{in}|$, is determined by the photon energy $E_\text{ph}$ and its incidence angle $\theta$:
\begin{equation}
    Q_z = \frac{4\pi\,E_\text{ph}}{h\,c_0}\sin{\theta}\ ,
    \label{eq:qz_A}
\end{equation}
where $h$ denotes the Planck constant and $c_0$ is the speed of light in vacuum.
Therefore, a scan of the reciprocal space can be obtained by either varying $\theta$, that changes the wave vectors' direction but keeping their magnitude fixed [see Fig.~\ref{fig:S0_qscan}(a)], or by varying $E_\text{ph}$, which changes the magnitude of $\Vec{k}\textsubscript{out}-\Vec{k}\textsubscript{in}$ [see Fig.~\ref{fig:S0_qscan}(b)]. 
Both methods provide the desired result, but employing white-light sources to common specular reflection geometries and adding an energy-dispersive detection increases the efficiency regarding acquisition time by enabling probing of large slices of reciprocal space in a single acquisition.
\begin{figure}[t!]
    \centering
    \includegraphics[width=1\columnwidth]{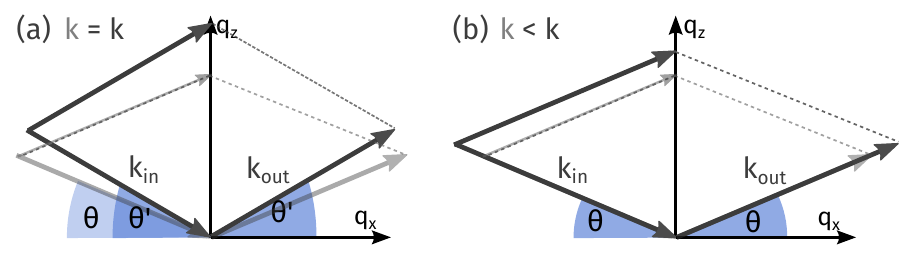}
    \caption{Scanning reciprocal space can be realized by varying the incidence angle (a) or the photon energy, which changes the length of the wave vector of the light (b).}
    \label{fig:S0_qscan}
\end{figure}
\begin{figure}[t!]
    \centering
    \includegraphics[width=1\columnwidth]{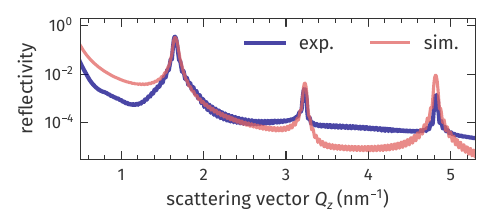}
    \caption{The simulated soft-x-ray scattering intensity as function of scattering vector $Q_z$ is compared to an experimental $Q_z$ scan measured in a monochromatic diffraction experiment at $E\textsubscript{ph}=1200\,$eV.}
    \label{fig:S1_qz_bessy}
\end{figure}

\section{Sample calibration via soft-x-ray simulations}

In this section, we describe the procedure of extracting the superlattice (SL) parameters by modeling the static Mo/Si SL soft-x-ray scattering response in reciprocal space and comparing it to an extended $Q_z$-scan measured in a soft-x-ray diffraction experiment at the PM3 beamline at BESSY~II.
The experimental data shown in Fig.~\ref{fig:S1_qz_bessy}(a) has been published previously~\cite{schi2016}, whereas the simulations are refined by a soft-x-ray scattering formalism~\cite{schi2021, elzo2012}.
The $Q_z$-scan has been obtained by varying the incident angle of the monochromatic soft-x-ray radiation at a fixed photon energy $E\textsubscript{ph}=1200\,eV$, which leads to a variation of the probed scattering vector $Q_z$, c.f. Eq.~(\ref{eq:qz}).
The probed $Q_z$-interval displays three distinct SL Bragg peaks of the order $n=1, 2, 3$. 
In the modeling, we first determined the thickness of the Mo/Si double-layer $d\textsubscript{DL}$, that defines the reciprocal lattice vector $|\Vec{G}\textsubscript{SL}| = 2\pi/d\textsubscript{DL}$ and therefore the position of the SL Bragg peaks in reciprocal space. The intensity peaks of consecutive order are separated by $\Vec{G}\textsubscript{SL}$, which allows for a precise determination of the real space DL thickness.
In a second step, we refined the sublayer thicknesses $d_i$ of the constituent SL materials Mo and Si. 
The ratio $d\textsubscript{Mo}/d\textsubscript{Si}$ can be understood as the SL form factor, which determines the scattered intensity amplitude for each SL Bragg peak of the order $n$. 
We obtained the best result using the parameters summarized in the top part of Table~\ref{tab:tab_1_sim_param} providing a reasonable agreement of the experimental and simulated data for the first and second SL Bragg peak.
As discussed in Ref.~\cite{schi2016}, deviations between the measured and simulated peak intensities, that occur for higher order SL peaks, can be attributed to thin inter-diffusion layers of $\text{MoSi}_2$ between the Mo and Si layers in the SL structure.

\begin{table*}[t!]
\centering
\begin{tabular}{l c c c }
 & Mo & Si SL & Si substrate  \\
 \hline
thickness $d$ (nm) & 1.88 & 2.05 & 615.00 \\
density $\rho$ (g\,cm\textsuperscript{-3}) & 9.252 & 2.336 & 2.336\\
\hline
heat capacity $C$ (J\,kg\textsuperscript{-1}\,K\textsuperscript{-1}) & 250 \cite{chou1984}& 712 \cite{abe2011} & 712 \\
therm. conductivity $\kappa$ (W\,m\textsuperscript{-1}\,K\textsuperscript{-1}) & 139 \cite{wen2020} & 1.8 \cite{quee2013} & 149 \cite{ashe1998}\\
corrected lin. therm. expansion $\alpha$ ($10^{-6}$ K\textsuperscript{-1}) & 9.86 \cite{whit1978, dick1967} & 4.25 \cite{lima1999, hopc2010} & 7.08 \cite{lima1999, hopc2010}\\
sound velocity $v_s$ (nm\,ps\textsuperscript{-1}) & 6.19 & 8.15 \cite{dech2014} & 9.4 \\
refractive index $n$\,(2.1\,\textmu m) & $3.3239+19.239j$ \cite{wern2009}& $3.4492+0j$  \cite{pier1972} & $3.4478+0j$ \cite{li1980}\\
\end{tabular}

\caption{\textbf{Thermo-elastic parameters of the Mo/Si superlattice}}
\label{tab:tab_1_sim_param}
\end{table*}

\section{Ultrafast SL strain response upon laser excitation}
\begin{figure*}[t!]
\centering
\includegraphics[width = \textwidth]{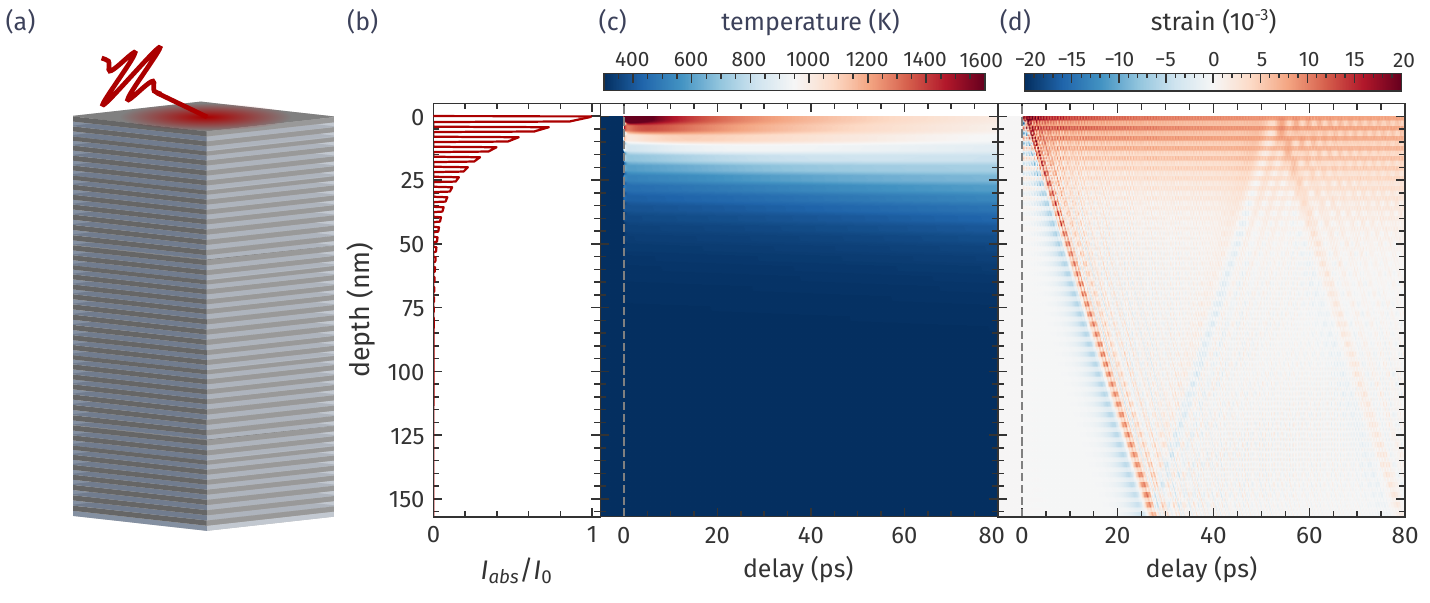}
    \caption{(a) The Mo/Si SL is built of alternating layers of Mo and Si. 
    (b) The metallic Mo mainly absorbs the IR pump energy resulting in an ultrafast heating of the metal layers. 
    (c) The modeled spatio-temporal temperature map shows ultrafast heating of the excited SL layers and a slow heat transfer into the sample depth. 
    (d) The rapid temperature increase induces a complex spatio-temporal strain response, that is composed of a quasi-static expansion and propagating strain waves.}
    \label{fig:S2_uf_dyn}
\end{figure*}
In this section, we briefly describe the modeling of the laser-induced change of the SL diffraction signal upon laser excitation in a purely one-dimensional treatment of the sample using the udkm1Dsim toolbox~\cite{schi2021} and the layer-specific parameters given in Table~\ref{tab:tab_1_sim_param}. 
The ultrafast excitation of the SL induces an ultrafast strain response of the lattice, leading to the temporal changes of the soft-x-ray scattering signal, which can be captured by the following steps:
First, we model the initial energy deposition by the 2.1\,\textmu m pump laser pulse according to a multi-layer matrix formalism.
As Si is transparent for the IR pulses used in the experiment, Mo is essentially absorbing the excitation light [see Fig.~\ref{fig:S2_uf_dyn}(b)].
From this initial depth-dependent energy distribution, we calculate the spatio-temporal temperature map shown in Fig.~\ref{fig:S2_uf_dyn}(c) using a Fourier-heat-diffusion model.
Due to the large band gap in Si and a fast electron-phonon coupling, that enables a fast equilibration of electrons and phonons in Mo \cite{well1999}, we can describe the temperature dynamics by a single temperature corresponding to the phonon system of each material. 
In the next step, we use these time- and depth-dependent temperatures to simulate the ultrafast strain response by numerically solving a linear-chain model of masses and springs.
We consider a Poisson-corrected linear thermal expansion coefficient, i.e. the ultrafast expansion coefficient, for both materials. 
The thermal expansion coefficient as well as the Poisson-factor of Mo are highly temperature-dependent, which is roughly taken into account by increasing both parameters in the model by an offset.
The complex spatio-temporal strain map shown in Fig.~\ref{fig:S2_uf_dyn}(d) is composed of a quasi-static expansion and coherently driven propagating strain pulses. 
We can identify the increasing background strain, that is of high amplitude near the surface according to the spatio-temporal temperature rise. 
Further, we observe a pronounced bipolar strain wave generated near the SL surface, that propagates towards the Si substrate interface, where it gets partially reflected after 25\,ps.
These structure dynamics are then translated to a time-dependent soft-x-ray scattering signal of the sample using the same x-ray scattering formalism as used for the static case. 
\newpage

\nocite{*}

\bibliography{20240708_manuscript}

\end{document}